# A Framework for Validation of Object-Oriented Design Metrics

Devpriya Soni[1]
*Department of Computer Applications*
*Maulana Azad National Institute of Technology (A Deemed University)*
*Bhopal 462007 India*
Ritu Shrivastava[2], M. Kumar[3]
*SIRT, Bhopal (India)*

***Abstract:*** **A large number of metrics have been proposed for the quality of object-oriented software. Many of these metrics have not been properly validated due to poor methods of validation and non acceptance of metrics on scientific grounds. In the literature, two types of validations namely internal (theoretical) and external (empirical) are recommended. In this study, the authors have used both theoretical as well as empirical validation for validating already proposed set of metrics for the five quality factors. These metrics were proposed by Kumar and Soni.**

***Keywords- object-oriented software, metrics, validation.***

## I. INTRODUCTION

Analyzing object-oriented software in order to evaluate its quality is becoming increasingly important as the paradigm continues to increase in popularity. A large number of software product metrics have been proposed in software engineering. While many of these metrics are based on good ideas about what is important to measure in software to capture its complexity, it is still necessary to systematically validate them. Recent software engineering literature has shown a concern for the quality of methods to validate software product metrics (e.g., see [1][2][3]). This concern is due to fact that: (i) common practices for the validation of software engineering metrics are not acceptable on scientific grounds, and (ii) valid measures are essential for effective software project management and sound empirical research. For example, Kitchenham et.al. [2] write: "Unless the software measurement community can agree on a valid, consistent, and comprehensive theory of measurement validation, we have no scientific basis for the discipline of software measurement, a situation potentially disastrous for both practice and research." Therefore, to have confidence in the utility of the many metrics those are proposed from research labs, it is crucial that they are validated.

The validation of software product metrics means convincingly demonstrating that:

1. The product metric measures what it purports to measure. For example, that a coupling metric is really measuring coupling.

2. The product metric is associated with some important external metric (such as measures of maintainability or reliability).

3. The product metric is an improvement over existing product metrics. An improvement can mean, for example, that it is easier to collect the metric or that it is a better predictor of faults.

According to Fenton [4], there are two types of validation that are recognized: internal and external. Internal validation is a theoretical exercise that ensures that the metric is a proper numerical characterization of the property it claims to measure. Demonstrating that a metric measures what it purports to measure is a form of theoretical validation. External validation involves empirically demonstrating points (2) and (3) above. Internal and external validations are also commonly referred to as theoretical and empirical validation respectively [2]. Both types of validation are necessary. Theoretical validation requires that the software engineering community reach a consensus on what are the properties for common software maintainability metrics for object-oriented design. Software organizations can use validated product metrics in at least three ways: to identify high risk software components early, to construct design and programming guidelines, and to make system level predictions. The approaches used in two validations are shown in Figure 1.

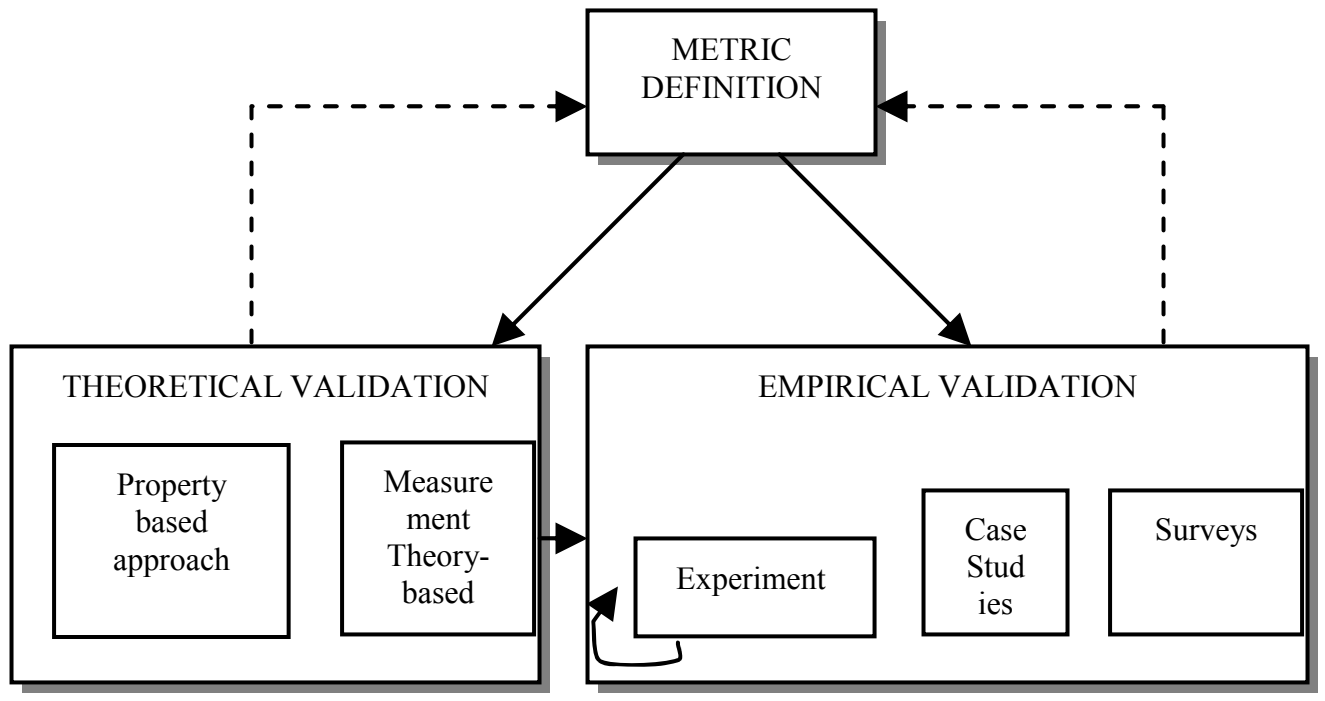

Figure 1 Approaches to software metrics validation

Recently, Kumar and Soni [5] have proposed a hierarchical model to evaluate quality of object-oriented software. This proposed model has been used for evaluation of maintainability assessment of object-oriented design quality, especially in design phase, by Soni and Kumar [6]. In this paper, the authors have attempted to validate the hierarchical model of object-oriented design quality metrics as given in [5]. The section II deals with theoretical validation of the model and the section III deals with empirical validation.

## II. Theoretical Validation of Proposed Hierarchical Model of Metrics

The main goal of theoretical validation is to assess whether a metric actually measures what it purports to measure [7]. In the context of an empirical study, the theoretical validation of metrics establishes their construct validity, i.e. it 'proves' that they are valid measures for the constructs that are used as variables in the study. There is not yet a standard, accepted way of theoretically validating software metric. Work on theoretical validation has followed two paths (see Fig 1):

- Measurment-theory based approach such as those proposed by Whitmire[8], Zuse[9], and Poels and Dedene [10]
- Property-based approach (also called axiomatic approaches), such as proposed by Weyuker and Braind et al.[11]

For the theoretical validation DISTANCE framework proposed by Poels and Dedene[9], is a conceptual framework for software metric validation grounded in measurement theory. This is briefly described in the next section.

### A. *The DISTANCE Measure Construction Procedure*

The measure construction procedure prescribes five activities. The procedure is triggered by a request to construct a measure for a property that characterizes the element of some set of objects. The activities of the DISTANCE procedure are given below. For notational convenience, let P be a set of objects that are characterized by some property pty for which a measure needs to be constructed.

*1) Finding a measurement abstraction:*The object of interest must be modeled in such a way that the property for which a measure is needed is emphasized. A suitable representation, called measurement abstraction hereafter, should allow to what extent an object is characterized by the property to be observed. By comparing measurement abstraction we should be able to tell whether an object is more, equally or less characterized by the property than other object.

*2) Defining distance between measurement abstraction:* This activity is based on a generic definition of distance that hold for elements in a set. To define distance between elements in a set, the concept of 'elementary transformation function' is used.

*3) Quantifying distance between measurement abstraction:* This activity requires the definition of a distance measure for the element of M. Basically this means that the distance defined in the previous activity are now quantified by representing i.e. measuring them as the number of elementary transformation by representing i.e. measuring them as the number of elementary transformations in the shortest sequence of elementary transformation between elements. Formally, the activity results in the definition of a metric $MxM \rightarrow R$ that can be used to map the distance between a pair of elements in M to a real number.

*4) Finding a reference abstraction:* This activity require a kind of thought experiment. We need to determine what the measurement abstraction for the object in P would look like if they were characterized by the theoretical lowest amount pty. If such a hypothetical measurement abstraction can be found, then this object is called the reference abstraction for P with respect to pty.

*5) Defining a measure for the property:* The final activity consists of defining a measure for pty. Since properties are formally defined as distances, and these distances are quantified with a metric function, the formal outcome of this activity is the definition of a function $\mu: P \rightarrow R$ such that $p \in P$: $\mu(p) = \delta(abs(p), ref(p))$.

### B. *Metric Validation*

The proposed model of Kumar and Soni [5] is reproduced in Fig 2 for ready reference. We have used the five activities of DISTANCE measure procedure for metrics of the model and important metrics are summarized in Table 1

## III. Empirical Validation of the Proposed Metrics

We have seen that survey is also commonly used method to empirically validate defined metrics. To obtain the view of persons who have fair experience of the software design and development, a questionnaire was prepared to validate metrics defined in the Fig 2. The questionnaire used for views is given in the appendix A. The first and second column respectively contains metrics names and their definitions. The respondents were asked to solicit their opinion in the form of yes, no or partially depending upon the metric effects on the five main quality factors, namely *functionality*, *effectiveness*, *understandability*, *reusability* and *maintainability*. The questionnaire was sent generously to two groups of people, the professionals working in industry like Infosys, TCS, Wipro, Accenture and people from academic institutes. We received 52 responses of which nearly 70% are from industry professionals and the rest from academic institutes. The analysis of the responses is done using Excel 2007. The results are since significant at 95% confidence level, on the whole if represents the opinion fairly. The analysis is presented in the next section.

**1** Functionality
- **1.1** Design Size
  - **1.1.1** *Number of Classes (NOC)*
- **1.2** Hierarchies
  - **1.2.1** *Number of Hierarchies (NOH)*
- **1.3** Cohesion
  - **1.3.1** *Cohesion Among Methods of Class (CAM)*
- **1.4** Polymorphism
  - **1.4.1** *Number of Polymorphic Methods (NOP)*
- **1.5** Messaging
  - **1.5.1** *Class Interface Size (CIS)*

**2** Effectiveness
- **2.1** Abstraction
  - **2.1.1** *Number of Ancestors (NOA)*
  - **2.1.2** *Number of Hierarchies (NOH)*
  - **2.1.3** *Maximum number of Depth of Inheritance (MDIT)*
- **2.2** Encapsulation
  - **2.2.1** *Data Access Ratio (DAR)*
- **2.3** Composition
  - **2.3.1** *Number of aggregation relationships (NAR)*
  - **2.3.2** *Number of aggregation hierarchies (NAH)*
- **2.4** Inheritance
  - **2.4.1** *Functional Abstraction (FA)*
- **2.5** Polymorphism
  - **2.5.1** *Number of Polymorphic Methods (NOP)*

**3** Understandability
- **3.1** Encapsulation
  - **3.1.1** *Data Access Ratio (DAR)*
- **3.2** Cohesion
  - **3.2.1** *Cohesion Among Methods of Class (CAM)*
- **3.3** Inheritance
  - **3.3.1** *Functional Abstraction (FA)*
- **3.4** Polymorphism
  - **3.4.1** *Number of Polymorphic Methods (NOP)*

**4** Reusability
- **4.1** Design Size
  - **4.1.1** *Number of Classes (NOC)*
- **4.2** Coupling
  - **4.2.1** *Direct Class Coupling (DCC)*
- **4.3** Cohesion
  - **4.3.1** *Cohesion Among Methods of Class (CAM)*
- **4.4** Messaging
  - **4.4.1** *Class Interface Size (CIS)*

**5** Maintainability
- **5.1** Design Size
  - **5.1.1** *Number of Classes (NOC)*
- **5.2** Hierarchies
  - **5.2.1** *Number of Hierarchies (NOH)*
- **5.3** Abstraction
  - **5.3.1** *Number of Ancestors (NOA)*
- **5.4** Encapsulation
  - **5.4.1** *Data Access Ratio (DAR)*
- **5.5** Coupling
  - **5.5.1** *Direct Class Coupling (DCC)*
  - **5.5.2** *Number of Methods (NOM)*
- **5.6** Composition
  - **5.6.1** *Number of aggregation relationships (NAR)*
  - **5.6.2** *Number of aggregation hierarchies (NAH)*
- **5.7** Polymorphism
  - **5.7.1** *Number of Polymorphic Methods (NOP)*
- **5.8** Documentation
  - **5.8.1** *Extent of Documentation (EOD)*

Figure 2 Proposed hierarchical design quality model

TABLE I. DISTANCE BASED VALIDATION CRITERIA FOR METRICS

| Quality Attribute | Metrics | Validation criteria | | | | |
|---|---|---|---|---|---|---|
| Functionality | | Measurement Abstraction | Defining distance between two extreme abstractions | Quantifying Distance in extremes. | Hypothetical reference abstraction | Defining a measure for pty |
| | | Object is more, equally or less characterized by the property than another object. | A set Te of elementary transformation function, sufficient to change any element of M into any other element of M. | M x M→R to map distance between a pair of elements in M to a real number. | Reference abstraction as a reference point for measurement. | μ:P → R such that p€ P:μ(p)= δ (abs(p), ref(p)) |
| | Number of Classes (NOC) | Total number of classes in the design | Various Classes available in the design | EQ={1,.8,.6,.4,.2,0} | EQ=1 if 8 or more classes | EQ=0 if no classes |
| | Number of Hierarchies (NOH) | Number of class hierarchies in the design | Various Classes available in the design | EQ={1,.8,.6,.4,.2,0} | EQ=1 hierarchy level is 5 or more | EQ=0 if no hierarchy |
| Effectiveness | Number of Ancestors (NOA) | Number of classes along all paths from the root class (es) to all classes in an inheritance. | Various Classes available in the design | EQ={1,.8,.6,.4,.2,0} | EQ=1 if 6 or more ancestors | EQ=0 if no ancestors |
| | Maximum Depth of Inheritance (MDIT) | Longest path from the class to the root of the hierarchy. | Various Classes in the hierarchy | EQ={1,.8,.6,.4,.2,0} | EQ=1 depth is 6 level or more | EQ=0 if depth is 1 level |
| | Number of Aggregation Hierarchies (NAH) | Total number of aggregation hierarchies. | Various classes/ objects/attributes | EQ={1,.8,.6,.4,.2,0} | EQ=1 if aggregation hierarchy 5 or more | EQ=0 if no aggregation hierarchy |
| Understandability | Cohesion Among Methods of Class (CAM) | Summation of the intersection of parameter of a method with the maximum independent set of all parameter types in the class. | Class/methods/ parameters | EQ={1,.8,.6,.4,.2,0} | EQ=1 if cohesion is between 5 or more classes | EQ=0 if no cohesion among methods |
| | Number of Polymorphic Methods (NOP) | Total methods exhibiting polymorphic behavior. | Classes/methods | EQ={1,.8,.6,.4,.2,0} | EQ=1 if methods with polymorphic behavior 5 or more | EQ=0 if no methods with polymorphic behavior |
| | Data Access Ratio (DAR) | Ratio of the number of private (protected) attributes to the total number of attributes declared in the class. | Private/Protected attributes and total attributes. | EQ={1,.8,.6,.4,.2,0} | EQ=1 if ratio is 80% or more | EQ=0 if ratio is less than 5% |
| | Functional Abstraction (FA) | Ratio of the number of methods inherited by a class to the total number of methods accessible by member methods of the class. | Classes/methods | EQ={1,.5,0} | EQ=1 if ratio is 80% or more | EQ=0 if ratio is less than 5% |
| Reusability | Direct Class Coupling (DCC) | Count of classes that are directly related by attribute declarations and message passing (parameters) in methods. | methods/parameters passing mechanism | EQ={1,.5,0} | EQ=1 if message passing is upto 5 or more classes | EQ=0 if no. of classes is 1 or less |
| | Class Interface Size (CIS) | Number of public methods in a class. | Input / output parameter | EQ={1,.5,0} | EQ=1 if public methods present are more than 5 | EQ=0 if public method absent |
| Maintainability | Number of Methods (NOM) | Number of methods defined in a class. | Classes/methods | EQ={1,.8,.6,.4,.2,0} | EQ=1 if methods per class are 6 or more | EQ=0 if no methods |
| | Number of Aggregation Relationship s (NAR) | Number of data declarations whose types are user-defined classes. | Various classes/ object attributes | EQ={1,.8,.6,.4,.2,0} | EQ=1 if number is more than 6 | EQ=0 if no aggregation relationship |
| | Extent of Documentati on (EOD) | Based on the documentation availability | Data dictionary present or not | EQ={1,.8,.6,.4,.2,0} | EQ=1 if documentation is upto 100% | EQ=0 if Documentation is upto 5% |

## A. Observations

*1) Number of Classes (NOC):* The Figure 3 illustrates that number of classes affects various quality factors in one way or other. 92.31% respondents agree that *functionality* gets affected by NOC. 90.38% have opinioned that *maintainability* gets affected by NOC and over 76.92% respondents agree that *reusability* gets affected by NOC.

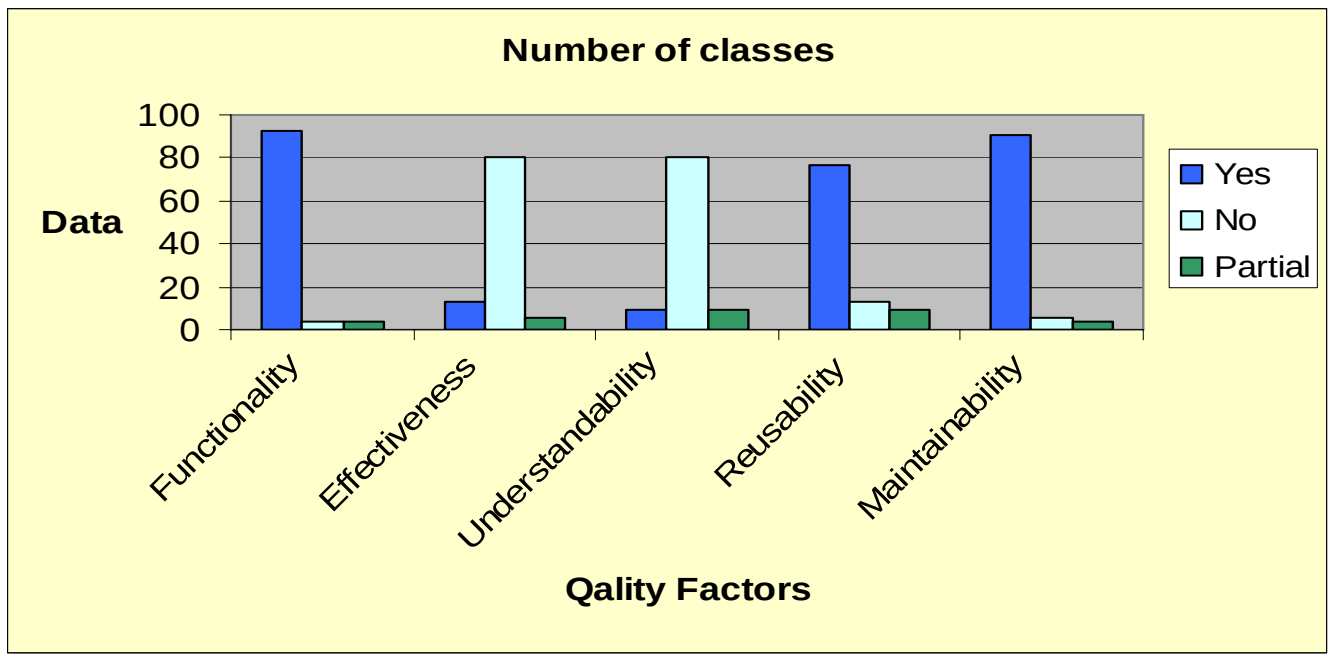

Figure 3 Impact of NOC on quality factors

*2) Number of Hierarchies (NOH):* The Figure 4 illustrates that number of hierarchies affects various quality factor in one way or other. 90.38% respondents agree that *functionality* gets affected by NOH. While 88.46% believed that *effectiveness* gets influenced by NOH. 78.85% have opinioned that *maintainability* gets affected by NOH.

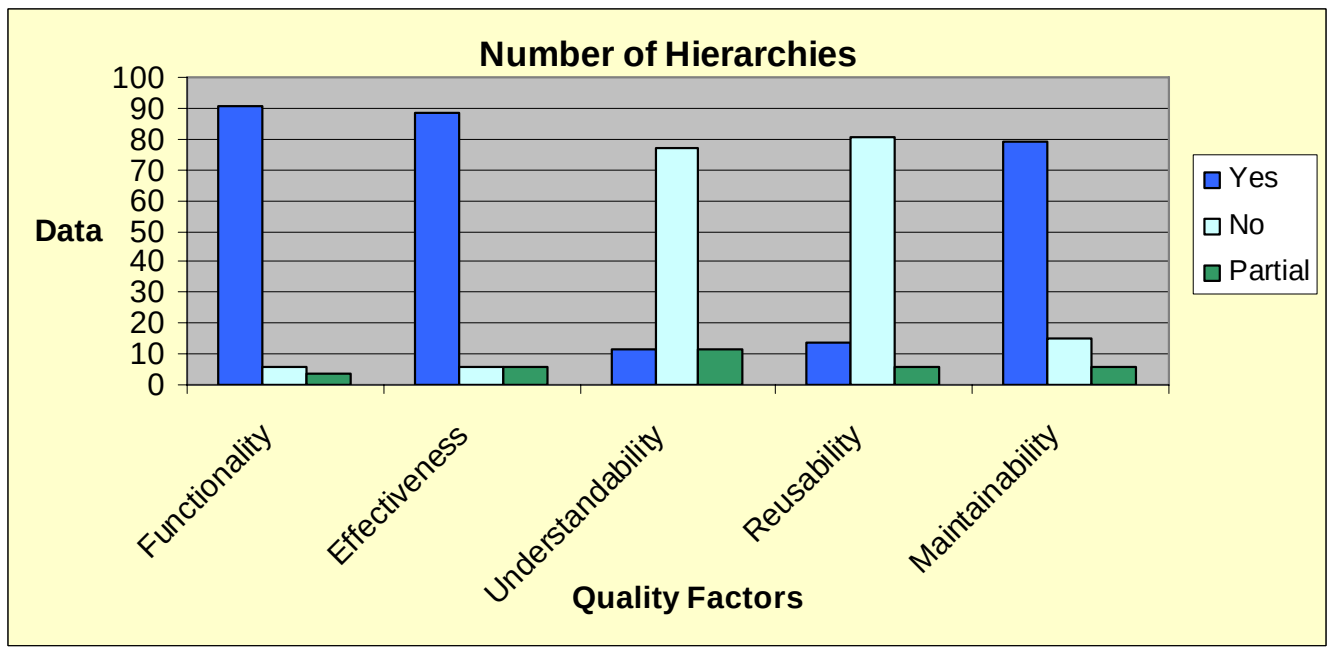

Figure 4 Impact of NOH on quality factors

*3) Cohesion Among Methods of Class (CAM):* 90.38% believed that *understandability* gets influenced by CAM. 84.62% have opinioned that *reusability* gets affected by CAM. 82.69% respondents agree that *functionality* gets affected by CAM.

*4) Number of Polymorphic Methods (NOP):* 86.54% respondents agree that *understandability* gets affected by NOP. 80.77% have opinioned that *functionality* gets affected by NOP. While 78.84% believed that *maintainability* gets influenced by NOP and over 75% respondents agree that *effectiveness* gets affected by NOP.

*5) Class Interface Size (CIS):* 90.38% respondents agree that *functionality* gets affected by CIS. While 82.69% believed that *reusability* gets influenced by CIS.

*6) Number of Ancestors (NOA):* 88.46% respondents agree that *effectiveness* gets affected by NOA. While 78.85% believed that *maintainability* gets influenced by NOA.

*7) Maximum Depth of Inheritance (MDIT):* 90.39% respondents agree that *effectiveness* gets affected by MDIT.

*8) Data Access Ratio (DAR):* 86.54% believed that *understandability* gets influenced by DAR. While 84.62% respondents agree that *effectiveness* gets affected by DAR. and over 76.92% respondents agree that *maintainability* gets affected by DAR.

*9) Number of Aggregation Relationships (NAR):*84.62% respondents agree that *maintainability* gets affected by NAR. While 78.85% believed that *effectiveness* gets influenced by NAR.

*10) Number of Aggregation Hierarchies (NAH):* 82.69% respondents agree that *effectiveness* gets affected by NAH. While 80.77% believed that *maintainability* gets influenced by NAH.

*11) Functional Abstraction (FA):* 80.77% respondents agree that *understandability* gets affected by FA. While 78.85% believed that *effectiveness* gets influenced by FA.

*12) Direct Class Coupling (DCC):* 84.62% respondents agree that *reusability* gets affected by DCC. While 80.77% believed that maintainability gets influenced by DCC.

*13) Number of Methods (NOM):* 82.69% respondents agree that maintainability gets affected by NOM.

*14) Extent of Documentation (EOD):* 75% respondents agree that *maintainability* gets affected by EOD.

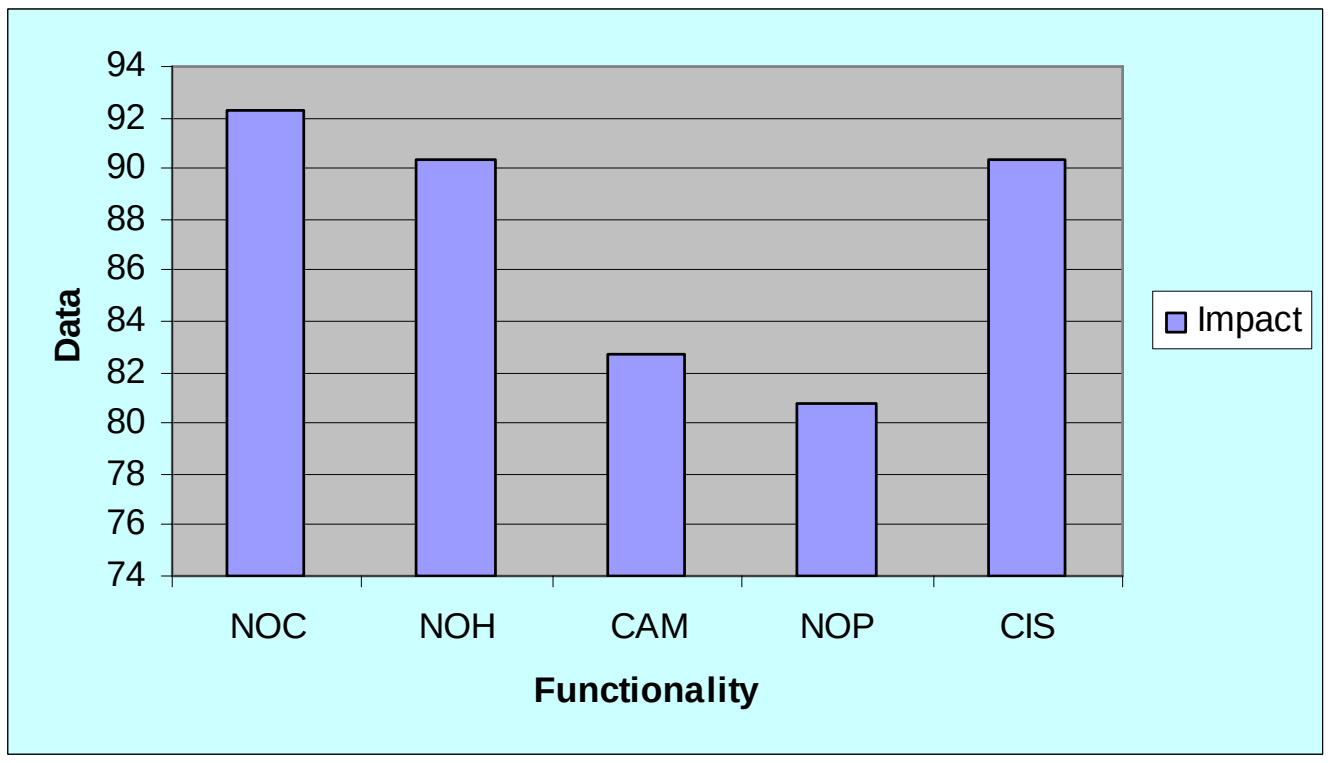

Figure 5 Impact of metrics on functionality

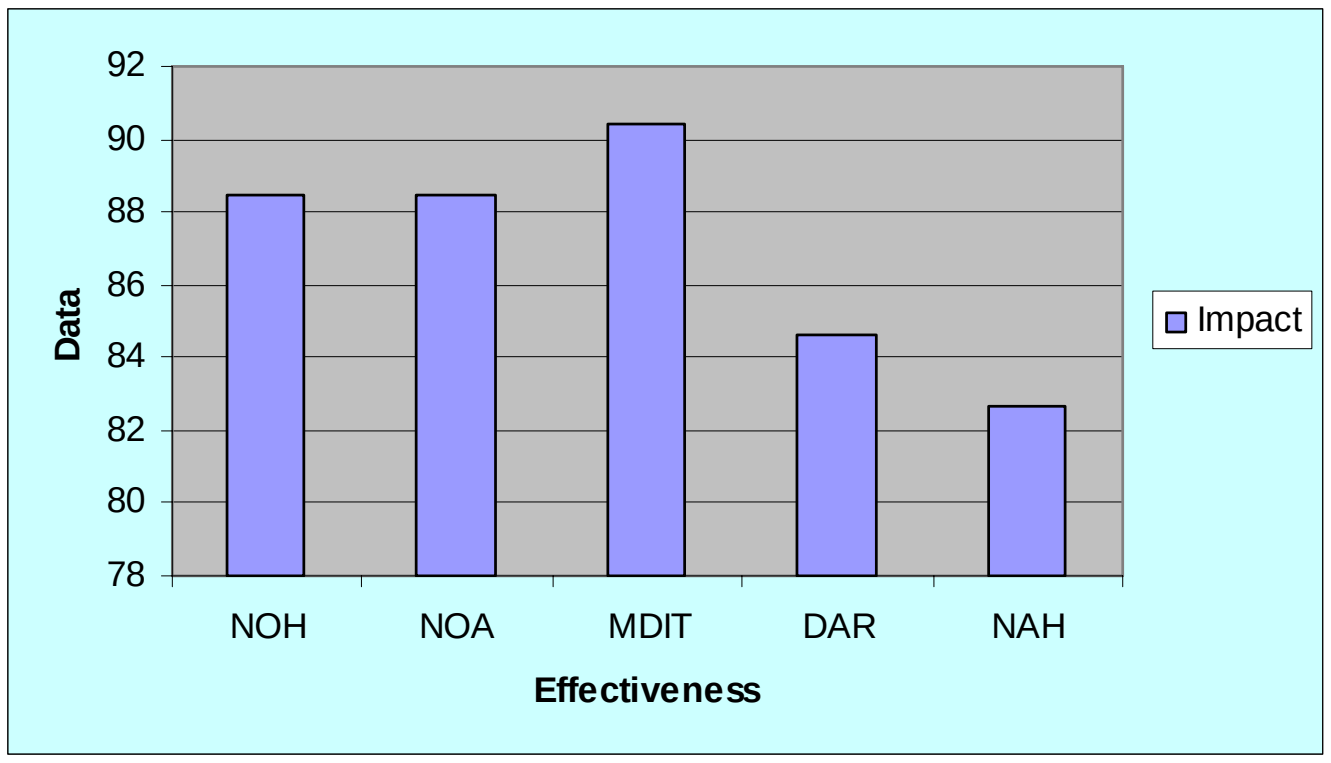

Figure 6 Impact of metrics on effectiveness

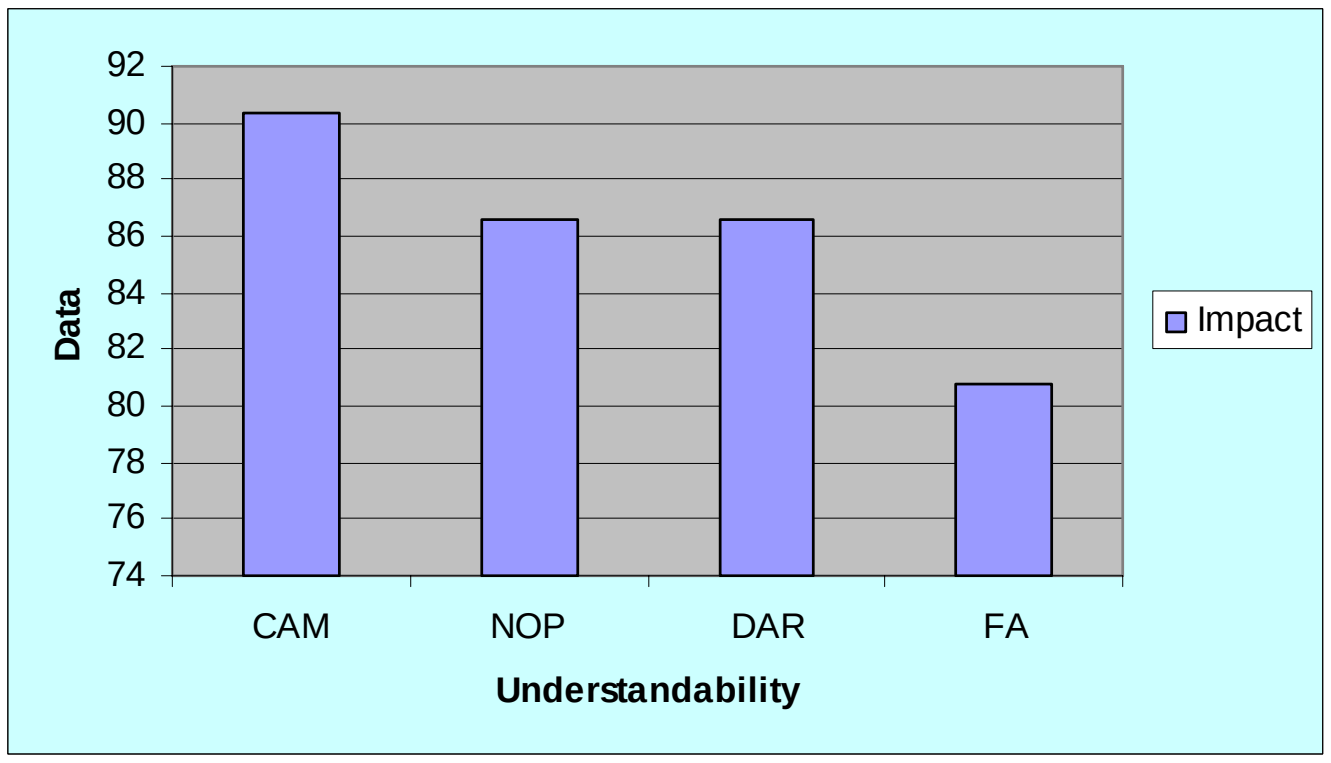

Figure 7 Impact of metrics on understandability

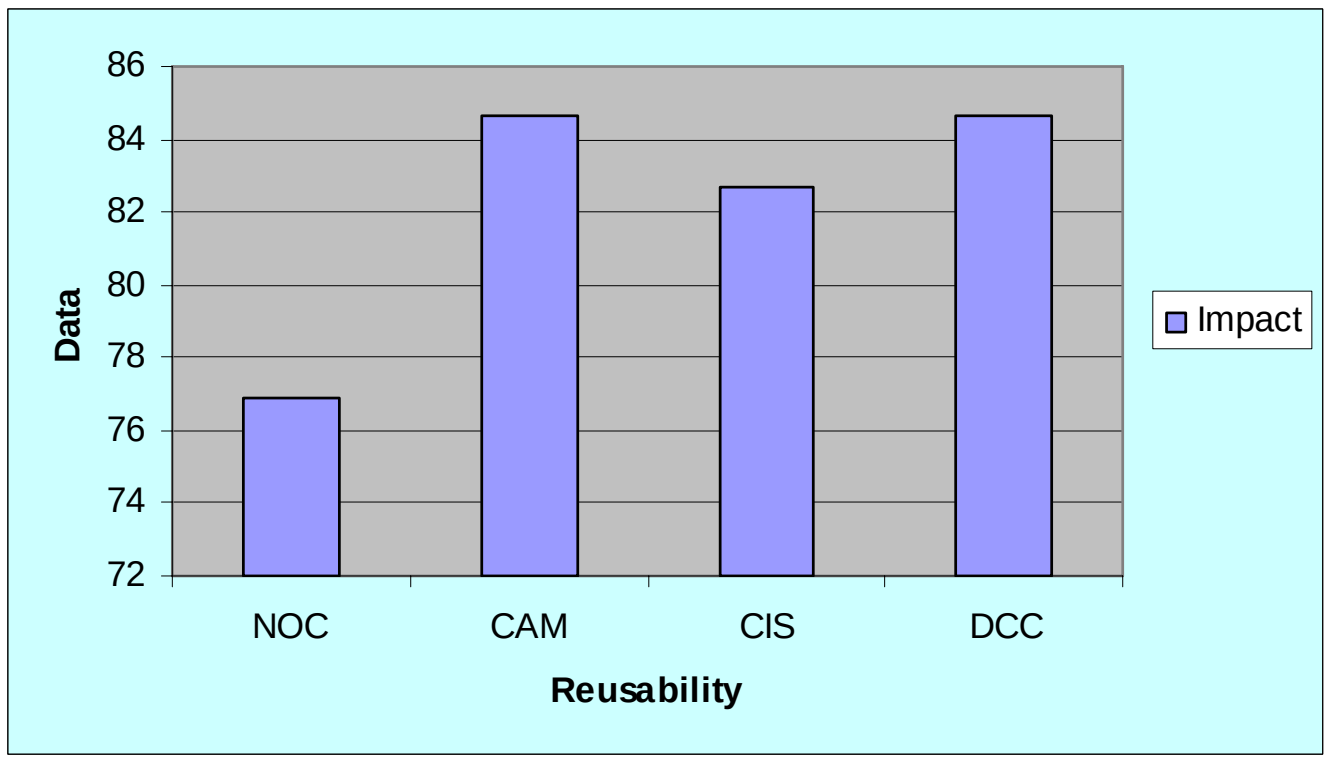

Figure 8 Impact of metrics on reusability

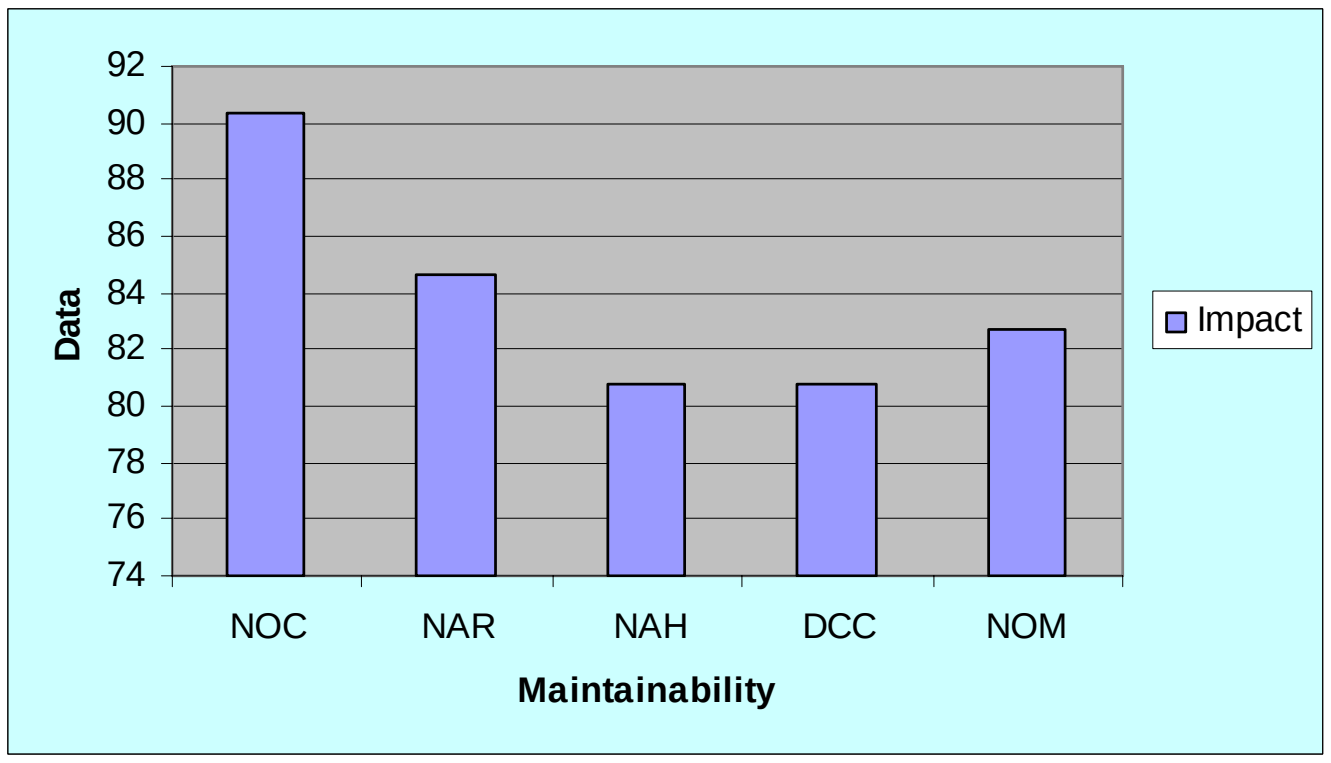

Figure 9 Impact of metrics on maintainability

## IV. CONCLUSION

A majority of respondents have opinioned that the metric NOC impacts three quality factors, *functionality*, *maintainability* and *reusability* and hence placement of NOC at these factors is justified. Further, majority respondents have opinioned that the metric NOH impacts three quality factors *functionality*, *effectiveness* and *maintainability* and hence placement of NOH at these factors is justified. Similar interpretations can be provided to other metrics. It is further observed that *functionality* is critically affected by the metric NOC followed by NOH (see Fig 5). *Effectiveness* is much affected by MDIT followed by NOH and NOA (see Fig 6). *Understandability* is much affected by CAM followed by NOP and DAR (see Fig 7). *Reusability* is much affected by CAM and DCC (see Fig 8). Similarly *maintainability* is much affected by NOC followed by NAR (see Fig 9). We have considered only five metrics in maintainability, however respondents opinioned that it is also affected by metrics NOH, NOP and NOA.

Appendix A

The following questionnaire was sent to respondents.

| Factor<br><br>Metric Name | Definitions | Functionality | Effectiveness | Understandability | Reusability | Maintainability |
|---|---|---|---|---|---|---|
| *Number of Classes (NOC)* | Total number of classes in the design | | | | | |
| *Number of Hierarchies (NOH)* | Number of class hierarchies in the design | | | | | |
| *Cohesion Among Methods of Class (CAM)* | Summation of the intersection of parameter of a method with the maximum independent set of all parameter types in the class. | | | | | |
| *Number of Polymorphic Methods (NOP)* | Total methods exhibiting polymorphic behavior. | | | | | |
| *Class Interface Size (CIS)* | Number of public methods in a class. | | | | | |
| *Number of Ancestors (NOA)* | Number of classes along all paths from the root class (es) to all classes in an inheritance. | | | | | |
| *Maximum Depth of Inheritance (MDIT)* | Longest path from the class to the root of the hierarchy. | | | | | |
| *Data Access Ratio(DAR)* | Ratio of the number of private (protected) attributes to the total number of attributes declared in the class. | | | | | |
| *Number of aggregation relationships (NAR)* | Number of data declarations whose types are user-defined classes. | | | | | |
| *Number of aggregation hierarchies (NAH)* | Total number of aggregation hierarchies. | | | | | |
| *Functional Abstraction (FA)* | Ratio of the number of methods inherited by a class to the total number of methods accessible by member methods of the class. | | | | | |
| *Direct Class Coupling (DCC)* | Count of classes that are directly related by attribute declarations and message passing (parameters) in methods. | | | | | |
| *Number of Methods (NOM)* | Number of methods defined in a class. | | | | | |
| *Extent of Documentation (EOD)* | Based on the documentation availability | | | | | |

AUTHORS PROFILE

**Devpriya Soni** has seven years of teaching experience to post graduate classes at MANIT and four years of research experience. She is pursuing her PhD at Department of Computer Applications, MANIT, Bhopal. Her research interest is object-oriented metrics and object-oriented databases.

**Ritu Shrivastava** has 12 years of teaching experience to graduate classes at MANIT and 2 years at Amity university at Delhi. She is pursuing research in object-oriented software engineering.

**Dr Mahendra Kumar** is presently Prof. & Dean of Computer Science at S I R T. Bhopal. He was Professor and Head Computer applications at M A N I T. Bhopal. He has 42 years of teaching and research experience. He has published more than 90 papers in National and International journals. He has written two books and guided 12 candidates for Ph D degree and 3 more are currently working. His current research interests are software engineering, cross language information retrieval, data mining, and knowledge management.